\documentclass{emulateapj}

\shorttitle{Large-Scale Disturbances with a Solar Eruption}
\shortauthors{Zong \& Dai}

\begin{document}

\title{CORONAL AND CHROMOSPHERIC SIGNATURES OF
LARGE-SCALE DISTURBANCES ASSOCIATED WITH A MAJOR SOLAR ERUPTION}

\author{Weiguo Zong\altaffilmark{1,3}, Yu Dai\altaffilmark{2,3}}
\affil{$^{1}$Key Laboratory of Space Weather, National Center for Space Weather, China Meteorological Administration, Beijing 100081, China}
\affil{$^{2}$School of Astronomy and Space Science, Nanjing University, Nanjing 210023, China}
\affil{$^{3}$Key Laboratory of Modern Astronomy and Astrophysics (Nanjing University),
Ministry of Education, Nanjing 210023, China}

\email{ydai@nju.edu.cn}

\begin{abstract}
We present  both coronal and chromospheric observations of large-scale disturbances associated with a major solar eruption on 2005 September 7. In GOES/SXI, arclike coronal brightenings are recorded propagating in the southern hemisphere. The SXI front shows an initially constant speed of 730 km s$^{-1}$ and  decelerates later on, and its center is near the central position angle of the associated coronal mass ejection (CME) but away from flare site. Chromospheric signatures of the disturbances are observed in both MLSO/PICS H$\alpha$ and MLSO/CHIP He I 10830 {\AA}, and can be divided into two parts. The southern signatures occur in regions where the SXI front sweeps over,  with the H$\alpha$ bright front coincident with the SXI front while the He I  dark front lagging the SXI front but showing a similar kinematics. Ahead of the path of the southern signatures, oscillations of a filament are observed. The northern signatures occur near the equator, with the H$\alpha$ and  He I fronts coincident with each other. They first propagate westward, and then deflect to the north  at the boundary of an equatorial coronal hole (CH). Based on these observational facts, we suggest that the global disturbances are associated with the CME lift-off, and show a hybrid nature: a mainly non-wave CME flank nature for the SXI signatures and the corresponding southern chromospheric signatures, and a shocked fast-mode coronal magnetohydrodynamics (MHD) wave nature for the northern chromospheric signatures.
\end{abstract}


\maketitle

\section{Introduction}
Solar eruptions, such as flares and coronal mass ejections (CMEs), often generate globally propagating disturbances in the solar atmosphere. In the chromosphere, large-scale disturbances were first discovered in H$\alpha$, which show as bright arclike fronts propagating away from the flare sites over distances in the order of $5\times10^5$ km with velocities ranging from 500 to 2000 km s$^{-1}$ \citep{moren60}. Such disturbances termed Moreton waves often induce oscillations of remote filaments \citep{smi71}. Since Moreton waves are too fast compared with the characteristic  velocity of a wave in the chromosphere (tens of km s$^{-1}$), \citet{uchi68} interpreted them as the skirt of a fast-mode coronal magnetohydrodynamics (MHD) wave that sweeps over the chromosphere. Observations of large-scale disturbances in the corona were becoming routinely available much later. With the launch of the Solar and Heliospheric Observatory (SOHO) spacecraft, quasi-circular diffuse coronal emission enhancements followed by expanding ``dimming regions'' are usually observed propagating outward from the eruption sites \citep{mos97,Thom99} by the onboard Extreme-Ultraviolet (EUV) Imaging Telescope \citep[EIT,][]{dela95}, and are therefore named ``EIT waves'' or ``EUV waves'' \citep{Thom00}.

In some EUV wave events the coincidence with a Moreton wave made it natural to take the EUV waves as a coronal fast-mode MHD wave \citep[etc.]{Thom00, poh01, warm01, warm05}, which serves as the coronal counterpart of the Moreton wave. Observations of filament oscillations induced by ``EUV waves'' supported the view \citep{take04}. However, there are still some challenges have to be overcome. In a statistical study, it was found that the velocities of EIT waves are 170-350 km s$^{-1}$ \citep{kla00}, much lower than those of Moreton waves. Furthermore, EUV waves often stop near the separatrix between active regions to form a stationary front \citep{dela99}. To reconcile the discrepancies, dual-wave models have been proposed. In these models, the first wave is a shock wave driven by the erupting flux rope with its leg sweeping the solar surface to form an H$\alpha$ Moreton wave, and the second wave observed as diffuse emission enhancement is not a true wave but magnetic reconfiguration in the framework of an expanding CME, which could be the density enhancement due to successive stretching of closed field lines \citep{chen02, chen05}, successive magnetic micro-reconnection between the expanding CME flank and low-lying loops nearby \citep{att07} or Joule heating in the current shell between the expanding flux tube and the surrounding fields \citep{dela08}.

Thanks to the technical improvements of the EUV Imagers \citep[EUVI,][]{how08} onboard the Solar Terrestrial Relations Observatory \citep[STEREO,][]{kai08} and the Atmospheric Imaging Assembly \citep[AIA,][]{lem12} onboard the Solar Dynamics Observatory (SDO), our knowledge on ``EUV waves'' is greatly improved. On one hand, some new observations indicate the wave nature of EUV wave, including reflection by a coronal hole (CH) or a coronal bright structure \citep{gop09, Li12, Ole12}, diffraction, refraction and reflection when the wave interacts with remote active region \citep{shen13}. On the other hand, two wave fronts are simultaneously observed in one event \citep{liu10,chen11}. The two fronts are usually indistinguishable during the CME strong lateral expansion, and their separation occurs when the CME lateral expansion slows down \citep{cheng12}. Afterwards, the fast one travels farther and indicates the fast-mode MHD wave nature until it fades out, while the slow one shows close correlation to the CME-caused restructuring \citep{liu12}, and trends to stop at a magnetic separatrix \citep{dai12}. These observations confirm the dual-wave models, in which an EUV wave front is initially composed of both wave and non-wave components before their separation.

Wavelike disturbances  have also been observed in soft X-ray (SXR) band \citep[etc.]{khan00, khan02, Naru02, hud03} and He I 10830 {\AA} filtergrams \citep[etc.]{vrs02, gil04}. Co-alignment analysis shows that signatures in these spectral channels are co-spatial and co-temporal with the Moreton waves and/or EUV waves, indicating that they are different signatures caused by a common disturbance and have similar behaviors \citep{warm05}. In this paper, we present observations on the large-scale disturbances associated with a major solar eruption in SXR band, H$\alpha$ line center and He I 10830 {\AA}. Different from previous studies, the signatures in SXR and H$\alpha$ (He I 10830{\AA}) show different shapes and locations, while kinematics analysis indicates they are from a common disturbance driven by the CME. The observations will help us better understand the nature of this kind of large-scale disturbances. In Section 2 we introduce the instruments and data sources. Observations are presented and analysis is carried out in Section 3. Then we discuss the results in Section 4 and draw our conclusions in Section 5.

\section{Data Sources}
The major solar eruption under study occurred on 2005 September 7 from NOAA active region (AR) 10808 when the  AR had just rotated onto the visible disk from behind. During the event, SOHO/EIT and SOHO/LASCO were not in operation. TRACE, which only provides partial disk images, was observing the Sun but pointing at another AR\@. Fortunately, the instruments located at Mauna Loa Solar Observatory \citep[MLSO, see http://mlso.hao.ucar.edu/mlso\_about.html or][]{tom05} were at work during this period. The initial evolution of the CME was recorded by the Mark-IV K-Coronameter (MK4). By rotating a pair of linear arrays in solar position angle around the center of the solar disk, MK4 scans the inner corona from 1.14 to 2.86 $R_\sun$ in a pixel resolution of about 11$^{\prime\prime}$. Observations of the disturbance signatures in the chromosphere were made by the Polarimeter for Inner Coronal Studies (PICS) in both limb and disk modes at the H$\alpha$ line center (6562.8 {\AA}) with a pixel resolution of 2.9$^{\prime\prime}$ and the Chromospheric Helium I Imaging Photometer (CHIP) at He I 10830 {\AA}  with a pixel resolution of 2.29$^{\prime\prime}$. CHIP also provides velocity images of the chromosphere. Both instruments recorded the data with a nominal time cadence of about three minutes.

Coronal signatures of the large-scale disturbances were recorded in images obtained by the Solar X-Ray Imager (SXI) onboard the Geostationary Operational Environmental Satellites (GOES) satellite. These images were acquired with the thin polymide filter (PTHN), which has a broad temperature response with a FWHM of  2 MK that peaks around 4 MK\@. SXI has a spatial resolution of about 5$^{\prime\prime}$ pixel$^{-1}$ and a high time cadence of $\sim$ 2 minutes. In addition, we checked metric radio dynamic spectrum to see radio signatures of the disturbances in the low corona. The spectrum was obtained by the Radio Solar Telescope Network (RSTN) in 25--180 MHz and the Green Bank Solar Radio Burst Spectrometer (GBSRBS, http://gbsrbs.nrao.edu/info.shtml) in $\sim$18--70 MHz, respectively.

\section{Observations and Data Analysis}
\subsection{Flare and CME}
Involved in the major solar eruption were an X-class flare and a fast CME\@.  The GOES 1-8 {\AA} SXR flux (shown in Figure 2) reveals that the flare starts at 17:17 UT\@, and peaks at 17:40 UT, reaching a level of X17. Considering the partial-occultation of the flare site by the limb, the actual peak flux should be higher.  During the event, PICS observed a flux rope rising from the active region and finally developing to a surge on the east limb. Figure 1 presents the evolution of the rising flux rope. The rising flux rope is first observed in the solar disk image at 17:23:34 UT; it looks like a bud extending out of the solar limb.  In the solar limb image it first appears at 17:26:59 UT\@. As shown in the following images, the light saturation prevents the main body of the rising flux rope from being recognizable until 17:36:04 UT\@. It is found that the flux rope is composed of many threads, which show an untwisting process during the eruption. After the threads are totally untwisted, the flux rope finally develops to a surge on the limb. In Figures 1(c) and (i) we also overlay the contours of  25--100 keV hard X-ray (HXR) emission reconstructed from the Reuven Ramaty High Energy Solar Spectroscopic Imager \citep[RHESSI,][]{lin02} data. It appears that the HXR source is located at the top of the flare site and north to the leg of the flux rope, with a position angle of S12.2$^{\circ}$.

\begin{figure*}
\epsscale{0.8}
\plotone{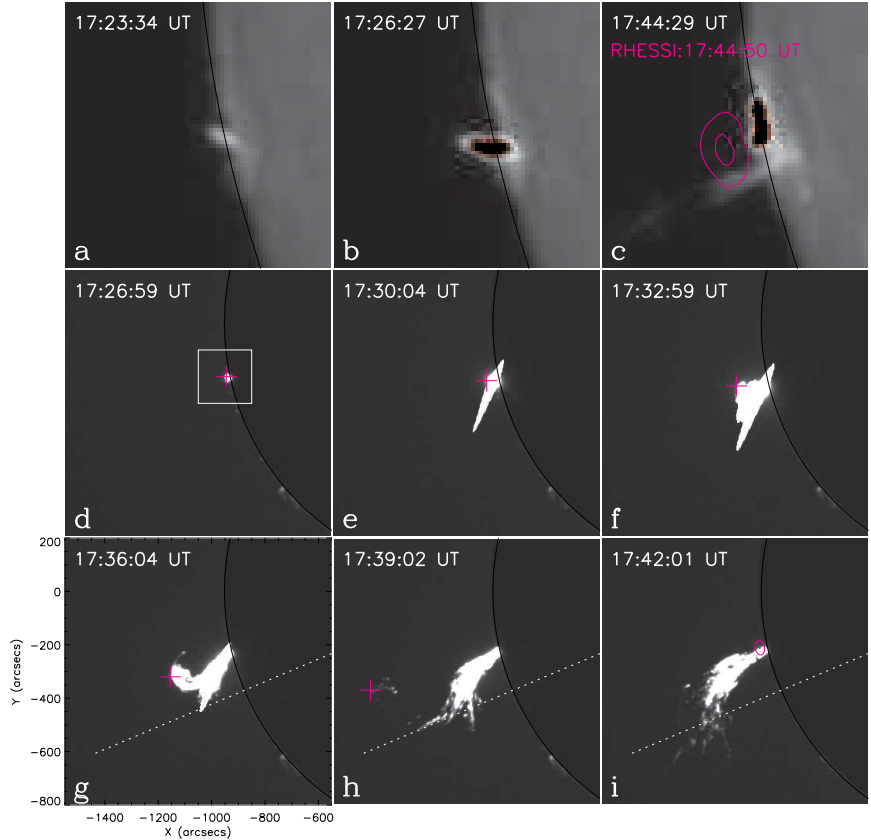}
\caption{PICS H$\alpha$ observations of the rising flux rope in disk (a-c) and limb (d-i) modes, respectively, among which the location and size of the disk images are outlined by the rectangle in panel (d). The top position of the flux rope is tracked by the red plus signs in panels (d-i). Overlaid in panels (c) and (i) is the RHESSI 25--100 keV HXR source at 17:44 UT in red contours.}
\end{figure*}

We determine the top position of the flux rope, which is indicated by the red plus signs in Figures 1(d-h), and plot its height versus time in Figure 2. By using a parabolic fit to the data points, it is found that the untwisting flux rope undergoes an acceleration of 1.44 km s$^{-2}$, and the outward speed of the flux rope reaches 1900 km s$^{-1}$ at 17:40 UT. Compared with the derivative of the SXR flux, which is generally taken as a proxy for the HXR emission \citep{new68}, the acceleration (or untwisting) process of the flux rope is consistent with the impulsive phase of the HXR emission.

\begin{figure}
\resizebox{\hsize}{!}{\includegraphics{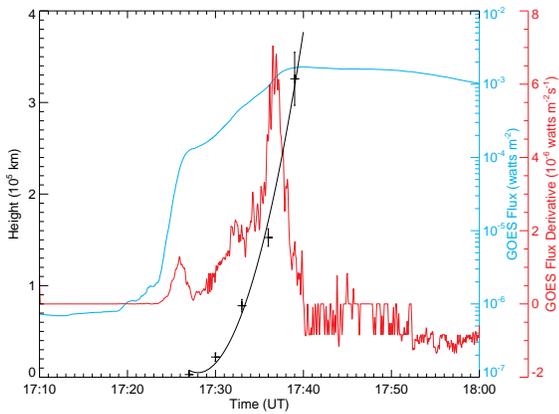}} \caption{
Height versus time diagram of the top position of flux rope (points with error bars) together with a parabolic fit. Also plotted are the time profiles of  GOES 1--8 {\AA} SXR flux and derivative of the SXR flux.}
\end{figure}

Figure 3 shows the coronal images obtained by MK4. As shown in Figure 3, a pre-existing helmet streamer is distorted by the rising flux rope, develops to an arc structure off the solar limb, and then erupts as a CME.  From 17:31 UT to 17:34 UT, the northern flank of the CME moves very little, while its southern flank moves about 15$^{\circ}$ in azimuth. Note that the time labeled in each frame refers to the starting moment of the corresponding scan. As a result of the asymmetric lateral expansion of the CME, the rising flux rope, which forms the bright core of the CME, is seen located north of the central position angle of the CME\@.  In Figure 3(c) we also co-align an SXI image near 17:40 UT, it is found that the southern flank of the CME is also visible in SXI.

\begin{figure*}
\epsscale{0.8}
\plotone{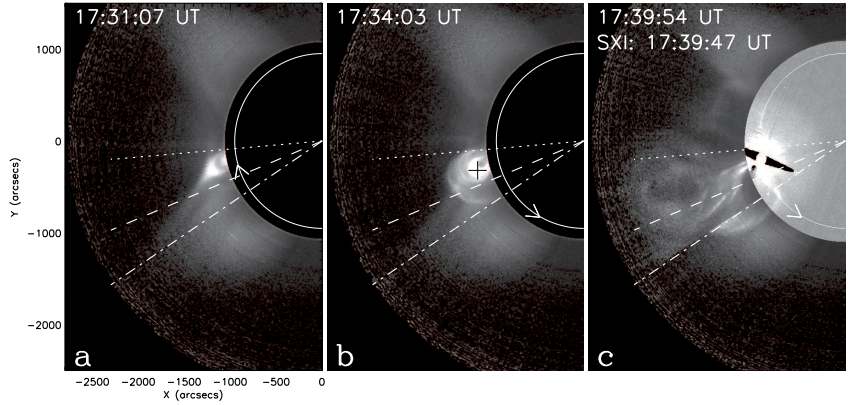}
\caption{MK4 observations of the inner corona, showing the evolution from a pre-existing helmet streamer (a) to an erupting CME (c). The dash-dotted line indicates the starting position of the linear sensor of MK4, and the arrows indicate the scanning directions. The dotted line depicts the initial azimuth angle of the CME's northern flank, and the dashed line extends across the fitted center of the SXI front (see the text). In panel (c) an SXI running difference image is co-aligned with the MK4 image. }
\end{figure*}

\subsection{SXI Wave}
Figure 4 shows running difference images obtained by SXI. Brightening signatures are first seen off the southeast limb, co-spatial with the southern flank of the CME. Then they propagate onto the solar disk as an arclike bright front. Finally the signatures fade out after they propagate over more than half the solar surface.

\begin{figure*}
\resizebox{\hsize}{!}{\includegraphics{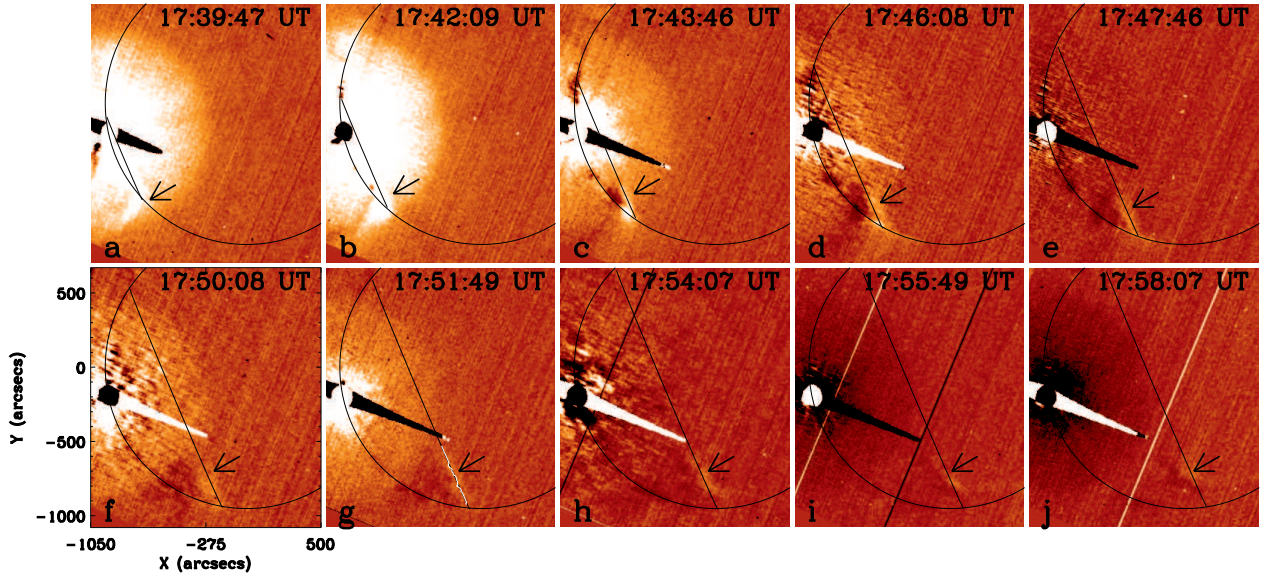}} \caption{SXI running difference images from 17:39:47 UT to 17:58:07 UT.  The arrows indicate the propagating SXI brightening front, and the black lines are the fitted projected circles for the brightening front by using the minimum $\chi^{2}$ method.}
\end{figure*}

To study the kinematics of the bright front in SXI, we use circles projected on the solar surface to forward fit the front assuming the circular wavefronts and isotropic propagation. In the first step, we try to find the center of the front. We choose the image at 17:51:49 UT (Figure 4(g)) for the reason that it shows the arclike front most clearly. We put a series of horizontal slices roughly perpendicularly across the front, and trace the intensity along each slice. The point of intensity maximum in each intensity profile is taken as the position of the front that intersects the corresponding slice. These positions are linked as a slightly zigzag white line shown in Figure 4(g).  By varying the position of the center of the circle with the longitude from -100$^{\circ}$ to -80$^{\circ}$ and the latitude from -10$^{\circ}$ to -30$^{\circ}$ in a step of 0.1$^{\circ}$, and the radius of the circle  from 0.8 to 0.9 $R_\sun$ in a step of 0.001 $R_\sun$, then we calculate the squared difference $\chi^{2}$ for each center position and radius that represents deviation of the fitted circle from the ``observed" front (the white line). The minimum $\chi^{2}$ corresponds to a center position at a latitude of -22.9$^{\circ}$ and a longitude of -90.9$^{\circ}$ (S22.9$^{\circ}$E90.9$^{\circ}$), and then this position is taken as the center of the front. In Figure 3(b), the radial direction of S22.9$^{\circ}$E90.9$^{\circ}$ is indicated by the dashed line, which is located near the central position angle of the the arc in MK4 (the CME in its initial phase in the low corona).

For the SXI images observed at other times, to fit the bright front, we use the same intensity profile method but fix the center position at S22.9$^{\circ}$E90.9$^{\circ}$. The fitted circles are plotted in Figure 4, with the project effect having been corrected. Note that they appears as straight lines because the center is very close to the limb.  By fitting the distance (radius of the fitted circles) versus time for the SXI front, which is presented in Figure 6(a), it is found that the SXI front propagates out with a constant velocity of 730 km s$^{-1}$ before 17:50 UT, and then shows a deceleration of 700 m s$^{-2}$ until it fades out in the SXI images.

The onset time of the SXI bright front is hard to be determined from the SXI images due to the light saturation near the flare site. By using the linear fit, the back-extrapolated onset time is about 17:35 UT, which is near the time when the untwisting process starts in the flux rope. However, we should bear in mind that the back-extrapolation assumes a point origin of the front. Therefore, the actual onset time should be somewhat later.

\subsection{Chromospheric Signatures}
Figure 5 shows PICS H$\alpha$ and CHIP He I 10830 {\AA} running difference images, both of which recorded global disturbances sweeping over the chromosphere. In H$\alpha$ the disturbances appear as bright features, while in He I 10830 {\AA} they are observed as dark features. According to their locations, the chromospheric disturbances can be divided into two parts. The southern chromospheric signatures (indicated by B$_\mathrm{S}$ and D$_\mathrm{S}$ in Figure 5) occur near the south limb, in regions where the SXI front sweeps over. They are faint but still can be recognized. The northern signatures (indicated by B$_\mathrm{N}$ and D$_\mathrm{N}$ in Figure 5) occur near the solar equator. They first propagate westward and then deflect into the north direction. Finally they disappear in the northern hemisphere.

\begin{figure*}
\epsscale{0.8}
\plotone{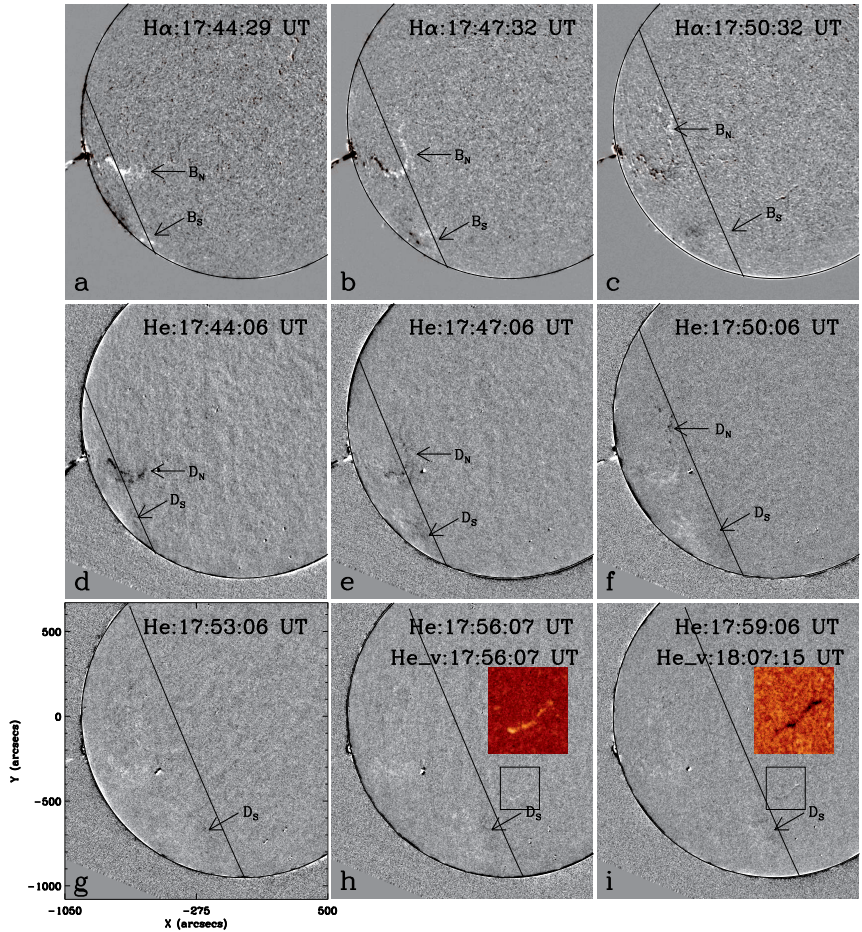}
\caption{PICS H$\alpha$ (a-c) and CHIP He I 10830 {\AA} (d-i) running difference images, showing the  disturbances sweeping over the chromosphere. The black lines indicate the position of the SXI front (interpolated before 17:50 UT or actually fitted after 17:50 UT) corresponding to the chromospheric observing time for each panel. The insets in panels (h and i) are the CHIP velocity maps highlighting the oscillation of a remote filament.}
\end{figure*}

We use the linearly fitted SXI front speed to interpolate the position of the coronal signatures corresponding to the chromospheric observing times before 17:50 UT, and plot it as projected circles in Figures 5(a-f). Since the the SXI front decelerates after 17:50 UT, in Figures 5(g-i), the black lines indicate the fitted projected circles for the SXI front observed at the nearest SXI observing time for each frame.

For the southern chromospheric signatures, it is seen that the positions of bright signatures in H$\alpha$ are coincident with those of the interpolated  coronal SXI disturbances, while the dark signatures in He I 10830 {\AA} lag the SXI ones. The H$\alpha$ brightenings fade out more quickly, barely discernible after 17:50 UT, while the dark front in He I 10830 {\AA} can be captured until 17:59 UT.  We also compare the kinematics of the southern chromospheric signatures to that of the SXI front. Due to the weak and diffuse nature of the southern chromospheric signatures, the intensity profile method, which is used to determine the positions of the SXI front, is unreliable for the chromospheric signatures. Therefore we alternatively use a visual track method. We outline a region of emission enhancement in H$\alpha$ (or emission depletion in  He I 10830 {\AA}) by point-and-click that encloses the most prominent chromospheric signatures, and calculate the centroid of the region. To minimize the subjectivity introduced by the visual selection, this procedure is repeated for several times, and the averaged position of the centroids is taken as the location of the corresponding chromospheric signatures. In fitting the distance versus time of the chromospheric signatures, we choose the same starting point (S22.9$^{\circ}$E90.9$^{\circ}$) as that for the SXI front. The fitting results are shown in Figure 6(a). As expected, the H$\alpha$ front propagates almost synchronously with the SXI front. As to the He I 10830 {\AA} front, although it lies behind the SXI front, with the distance between the two fronts ranging from 39 Mm to 63 Mm, it  shows a quite similar kinematics to that of the SXI front, with a slightly higher deceleration of 880 m s$^{-2}$  after 17:50 UT. At about 17:56 UT, a filament located at S30$^{\circ}$W04$^{\circ}$ (outlined by the black rectangle in Figures 5(h-i)), which is ahead of the inferred trajectory of the SXI front, starts to oscillate. The velocity images from CHIP recorded the oscillation process, during which the filament body shows redshift first (brightening in the zoom-out frame in Figure 5(h)), and about ten minutes later, it turns to blueshift (darkening in the zoom-out frame in Figure 5(i)). Simultaneously, the filament winks in H$\alpha$ images. It is seen that temporal and spatial position of the start of the filament oscillation (asterisk in Figure 6(a)) is very close to linear part extrapolation of the fitted SXI front kinematics.

\begin{figure}
\epsscale{0.95}
\plotone{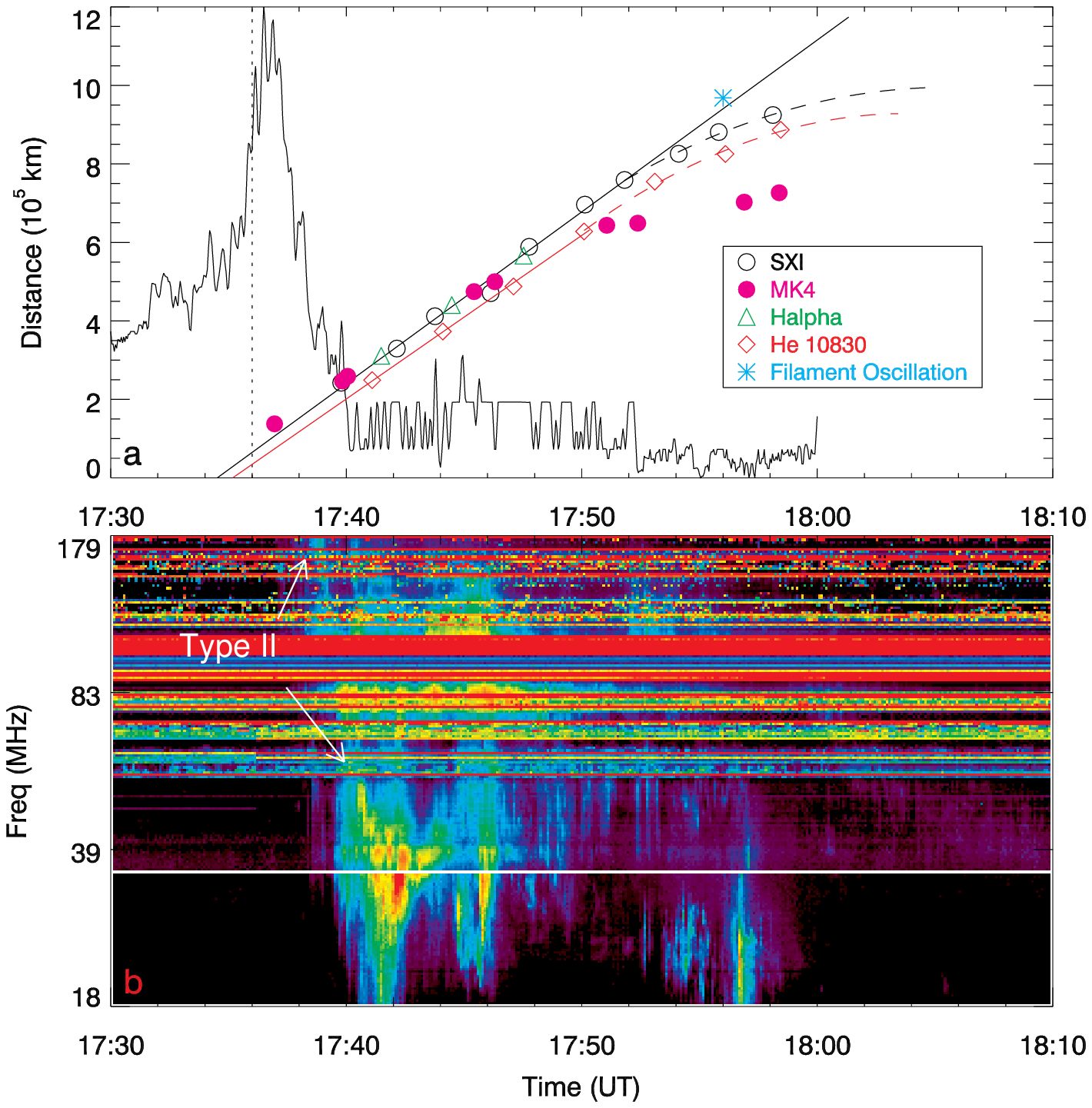} 
\caption{(a) Distance versus time diagrams of the CME's southern flank in MK4, the fitted SXI front, and the corresponding southern chromospheric signatures together with the kinematic fits to the data (linear before 17:50 UT and parabolic after 17:50).  Over-plotted is the temporal derivative of GOES SXR flux, which serves as a proxy for the HXR emission. The vertical dotted line represents the extrapolated onset time of the SXI disturbances. Meaning of the symbols is shown in the legend.  Note that the temporal and spatial position of the start of the filament oscillation is very close to linear part extrapolation of the fitted SXI front kinematics. (b) Metric radio dynamic spectrum combining the observations from RSTN (25--180 MHz) and GBSRBS (17--70 MHz). The arrows point to a metric type II burst occurring between 17:42 UT and 17:50 UT in the observing frequency range.}
\end{figure}

For the northern chromospheric signatures, they are first seen at 17:44 UT and can be recognized in base difference images until 17:59 UT\@. It is found that the signatures show similar morphology in both chromospheric channels, with the disturbance positions in the two chromospheric channels  coincident with each other.  Compared to the rather diffuse southern ones, the northern chromospheric signatures appear as a relatively sharp front. With the aid of the projected circles as distance baselines, it is clearly seen that the northern  signatures decelerate significantly in the west direction,  as they intersect with the projected circles at first (see Figures 5(a, b, d, e)) but lie behind later on (see Figures 5(c, f)). Meanwhile, the northern signatures deflect northward, turning the morphology to an elongated ``J'' shape. Due to the irregular trajectory of the northern signatures, it is not easy to quantitatively determine their kinematic parameters. However, it is rather interesting to address where they exactly deflect.  We extrapolate the coronal magnetic topology from the SOHO/Michelson Doppler Imager \citep[MDI,][]{sch95} synoptic magnetogram with the potential-field source-surface \citep[PFSS,][]{sch03} model, and overlay it on the He I 10830 {\AA} base difference image at 17:50 UT in Figure 7. Revealed by the magnetic field lines and further validated by the He I 10830 {\AA} plain images (not shown in this paper), there is an equatorial CH (from which read open field lines emanate) ahead of the propagation path of the northern signatures. The deflection of the northern signatures just occurs at the eastern boundary of the CH\@. We note that similar wave-CH interactions have been reported in some other works, including reflections of waves at CH boundaries in the corona \citep[eg,][]{gop09}, and stopping of a chromospheric Moreton wave at a CH boundary  \citep[cf.][]{Veronig06}.

\begin{figure}
\epsscale{0.8}
\plotone{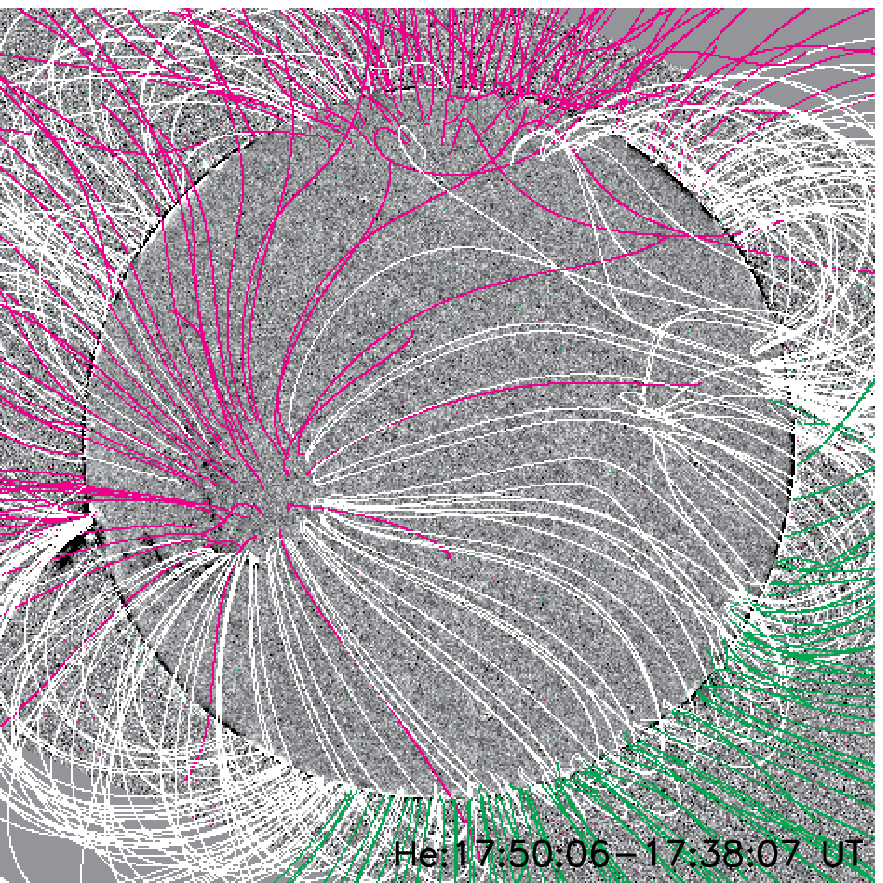}
 \caption{ Coronal magnetic topology extrapolated from the MDI synoptic magnetogram with the PFSS model overlaid on the He I 10830 {\AA} base difference image. The red and green lines represent open magnetic field lines from negative and positive polarities, respectively.}
\end{figure}
\section{Discussions}
\subsection{Initiation of the Disturbances}

Some authors propose that global disturbances such as ``EUV waves'' are generated by the pressure pulse of a solar flare \citep[e.g.,][]{wu01, vrs02}, while others suggest that the ``EUV waves'' are associated with CMEs \citep[e.g.,][]{dela99, chen02}. In this event, the observations are more likely tosupport the latter view. The fitted center of the SXI arclike bright front is located at S22.9$^{\circ}$E90.9$^{\circ}$, which is near the central position angle of the arc structure in Figure 3(b) but 10$^{\circ}$ apart from the HXR source of the flare  (see Figures 1(c and i)). Moreover, the CME's southern flank overlaps the SXI bright signatures off the solar east limb in Figure 3(c) in the early stage. To further validate this argument, we trace the lateral expansion of the CME along the solar limb. The position angle of the CME's southern flank is determined at the occulted limb of MK4, which is at a height of $\sim$1.14 $R_{\sun}$, and the observing time of each position angle is obtained with the {\tt mk4\_time.pro} procedure within the Solar Software (SSW). Since the center of the SXI front is very close to the limb, we can take the radial direction of this point as the reference position angle and convert position angle of the CME's southern flank to the distance along the solar limb. The evolution of the CME's lateral expansion is shown in Figure 6(a) for comparison. It is clearly seen that the CME's southern flank follows nearly the same kinematics as that of the SXI front before 17:50 UT. Afterwards the CME's lateral expansion also slows down but shows a deviation from the SXI front. The reason for this deviation will be further discussed in the following.

The back-extrapolated onset time of the SXI disturbances is about 17:35 UT, simultaneous (within 1 minute) with the peak of the derivative of the SXR flux, which serves as a proxy for the HXR emission (see Figure 6(a)). As mentioned above, the impulsive phase of the HXR emission corresponds to the fast acceleration (or untwisting) process of the flux rope, which forms the core of the CME. In this sense, the magnetic reconnection in the flare site, the lift-off the CME, and the onset of the SXI disturbances are closely correlated and probably involved in the same physical process.

\subsection{Nature of the Disturbances}
The SXI disturbances are initially co-spatial with the southern flank of the CME, and then propagate onto the solar disk, showing a correspondence to the southern chromospheric disturbances. Kinematic analysis of the SXI bright front reveals a constant velocity of 730 km s$^{-1}$ before 17:50 UT and a deceleration of 700 m s$^{-2}$ afterwards. Shortly after the SXI front slows down, a filament in front of the SXI front starts to oscillate. This process is very similar to that presented in \citet{dai12}. The SXI disturbances should mainly correspond to the CME flank. Lateral expansion of the CME will enhance SXR emission in this flank and leave footprints in the chromosphere. Meanwhile a fast-mode coronal MHD wave is driven ahead of the CME flank. We conjecture that the wave speed is very close to that of the CME's lateral expansion, 730 km s$^{-1}$ here, which could be indeed the case for a typical coronal environment. Therefore, this fast-mode wave remains attached on the CME flank until the CME's lateral expansion slows down. Afterwards, the fast-mode wave separates from the CME flank and travels freely with an unaltered speed, inducing oscillations of the ahead filament, as expected from the linear part extrapolation of the fitted SXI front kinematics. We note that this scenario is also nearly the same as that proposed by \citet{pat09} in explaining the 2009 February 13 EUV wave event, which was observed by the twin \emph{STEREO} spacecrafts in quadrature. In their event, while the widths of the CME (from limb view) and the EUV wavefront (from on-disk view) track each other quite closely in the beginning, the wavefront becomes significantly wider than the CME later on. They concluded that the EUV wave signatures purely correspond to a fast-mode MHD wave driven by the lateral expansion of the CME\@. It seems that the situation is very similar to what is seen from the kinematics of the CME's southern flank and the SXI disturbances in our event.  However, if the SXI bright front were a fast-mode MHD wave, it would induce the filament oscillations only when it propagates to the filament body, which is not the case in this event.  In this sense, we prefer that the southern chromospheric disturbances reflect a non-wave CME component. As to kinematic deviation of the CME's southern flank and the SXI front in the late stage, we alternatively attribute it to the fact that the kinematics of the CME's southern flank is traced along the limb but the SXI front has propagated onto the disk later on. Although we believe that the  SXI front (or CME)-driven fast-mode wave responsible for the filament oscillations does exist, in SXI we can not catch any signatures of this wave component. We attribute it partially to the weakness nature of the wave, which is further validated by the fact that there are no other signatures of this wave component in the chromospheric channels besides the filament oscillations, and partially to the low sensitivity of SXI.  Since the perturbation of the ``invisible" wave comes from the above, it will push the filament downward so that in the CHIP velocity maps the filament body shows redshift first.

Compared to the southern chromospheric disturbance signatures, the northern ones are relatively stronger and sharper. They decelerate when propagating westward, and then deflect to the north at the boundary of an equatorial CH\@. We also check the metric radio dynamic spectrum during the interval of interest. As shown in Figure 6(b), the most prominent feature in the dynamic spectrum is a metric type II burst between 17:42 UT and 17:50 UT that covers the whole observing frequency range (180 --18 MHz), reflecting the existence of a coronal shock. Note that the actual starting frequency (time) of the metric type II burst should be higher (earlier) since it is first seen at the upper limit of the RSTN frequency range. The interval of the type II burst is consistent with the period when the northern chromospheric signatures are the most prominent. Therefore, it is reasonable to link the sharp northern chromospheric signatures to the coronal shock. Just as proposed by \citet{uchi68} in explaining Moreton waves, the skirt of the coronal shock sweeps over and perturbs the dense chromosphere, leaving a sharp front seen in the chromospheric channels. In this sense, the northern chromospheric disturbances should reflect a shocked fast-mode MHD wave component. Nevertheless, in SXI there is not any coronal signatures corresponding to the northern chromospheric signatures. Why do the stronger northern chromospheric signatures (compared to the southern ones) lack discernible coronal counterpart in SXI\@? One possibility is that the westward lateral expansion of the CME, which should be responsible for the northern chromospheric signatures, may be very short and impulsive so that it would stop very close to the flare site. As a consequence of the absence of continuing driving agency in this direction, the westward propagating part of the shock decays in both velocity and intensity, as expected from the deceleration of the northern chromospheric signatures. Therefore, in SXI, the light saturation can easily hide the laterally expanding CME flank in the early stage, while the low sensitivity of SXI may prevent it from capturing the continuously decaying coronal shock later on.Finally, as expected from the wave scenario, the wave front is naturally reflected by the CH it encounters.

The He I 10830 {\AA} disturbances show as dark features, implying an enhancement of the absorption in He I 10830 {\AA} line. This absorption enhancement can be interpreted as either the collision process caused by a coronal shock \citep{vrs02}, or the photoionization-recombination (PR) process due to an increase of coronal radiation at wavelengths less than 504 {\AA} penetrating the upper chromosphere \citep{gil04}. Note that \citet{vrs02} actually consider also the PR process. In the latter view, the recombination process will take about 1 minute so that the He I 10830 {\AA} signatures will lag the above coronal emission enhancement by a distance. With the velocity of the SXI front (730 km s$^{-1}$) in this work taken into account, the lag distance is about 44 Mm, lying in the measured distance range (39--63 Mm) for the southern chromospheric disturbances. Although projection effects would also result in an apparent mismatch of chromospheric signatures and their coronal counterpart \citep[e.g.,][]{warm04,hud03}, in this event the spatial coincidence between the southern H$\alpha$ front and the SXI front indicates that the project effect, if existed, would be marginal. Therefore, it seems that the PR process plays a main role in the formation of the southern He I 10830 {\AA} dark signatures. To our knowledge, it may be the first report of a He I 10830 {\AA} front keeping following its coronal counterpart by lagging a distance. For the northern chromospheric disturbances, as we propose a shocked fast-mode coronal MHD wave in nature, there should be co-spatial He I 10830 {\AA} and H$\alpha$ signatures in the chromosphere.

\section{Summary and Conclusions}
Multi-wavelength analysis allows us to make a comprehensive study on the large-scale disturbances associated with the major event on 2005 September 7. Based on the results discussed above, we conclude that the global disturbances are associated with the CME lift-off, and show a hybrid nature: a mainly non-wave CME flank nature for the SXI signatures and the corresponding southern chromospheric signatures, and a shocked fast-mode coronal MHD wave nature for the northern chromospheric signatures. A new feature is found in this event: in the southern chromospheric disturbances, the He I 10830 {\AA} dark signatures lag the SXI/H$\alpha$ bright signatures by a distance, while in the northern chromospheric disturbances, the He I 10830 {\AA} dark and the H$\alpha$ bright signatures are co-spatial with each other.  This can be reasonably explained in the framework of the hybrid nature for the global disturbances.

\begin{acknowledgements}
We thank the anonymous referee whose constructive suggestions and comments significantly improve the manuscript. We are grateful to SOHO/EIT, SOHO/MDI, SOHO/LASCO, GOES, RHESSI and GBSRBS teams for their open data policy. This work is supported by NSFC (40904056, 11103009) and 973 project of China (2011CB811402, 2014CB744203) .
\end{acknowledgements}

\end{document}